\documentclass[a4paper,groupcitations]{article}

\usepackage[utf8]{inputenc}
\usepackage{lmodern}
\usepackage[english]{babel}
\usepackage{amsfonts, amsmath, amssymb, amsthm}
\usepackage{array}
\usepackage{graphicx}
\usepackage{subfigure}
\usepackage{braket}
\usepackage{eucal}
\usepackage{verbatim}
\usepackage[table]{xcolor}
\usepackage{caption}
\usepackage{cite}
\usepackage{textcomp}
\usepackage{ragged2e}
\usepackage[colorlinks=true,linkcolor=blue,citecolor=blue,urlcolor=blue]{hyperref}
\usepackage{tikz}
\usetikzlibrary{positioning,arrows,decorations.pathmorphing,decorations.markings,calc,shapes,matrix}
\usepackage{pgfplots}
\usepackage{subcaption}

\newcommand{\be}{\begin{eqnarray}}
	\newcommand{\ee}{\end{eqnarray}}

\def\theequation{\thesection.\arabic{equation}}

\title{\bf Entanglement Entropy of a Non-Minimally Coupled Self-Interacting Scalar across a Schwarzschild Horizon at $\mathcal{O}(\alpha)$}
\author{
	Florin Manea\thanks{florin.manea1@s.unibuc.ro} \\
	{\em Faculty of Mathematics and Computer Science, University of Bucharest} \\
	{\em Iuliu Maniu Boulevard 15, Bucharest}
}

\begin{document} 
	
	\maketitle
	
	\vspace{0.5cm}
	\noindent\textbf{\small Abstract:} \justifying
	We compute the first-order correction in the quartic coupling $\alpha$ to 
	the entanglement entropy of a massive, non-minimally coupled scalar across 
	the horizon of a four-dimensional Schwarzschild black hole, treating the 
	non-minimal coupling $\xi$ as a free parameter. Combining the replica 
	trick on the conical manifold $\mathcal{M}_n$ with heat-kernel methods in 
	proper-time regularization, we obtain a closed-form expression for 
	$\delta S^{(1)}(m,\alpha,\xi)$. The bare correction exhibits a 
	log-enhanced quadratic divergence $\epsilon^{-2}\ln(m^2\epsilon^2)$, 
	arising from interference between bulk fluctuations and the distributional 
	curvature at the tip; we show it is cancelled at $\mathcal{O}(\alpha)$ by 
	the bulk mass counterterm. The residual $m^2\ln(m^2\epsilon^2)$ divergence 
	renormalizes Newton's constant, preserving $S_{\mathrm{BH}} = 
	\mathcal{A}_\Sigma / 4 G_F$. The correction is proportional to $(1/6-\xi)$ 
	and vanishes identically for conformal coupling.
	\vspace{0.5cm}

	\noindent\textbf{\small Keywords:} black hole entropy; entanglement entropy; replica trick; heat kernel; scalar field self-interactions;
	Newton's constant renormalization
	
\section{Introduction}
	\hspace{1em} Black hole entropy, first proposed by Bekenstein~\cite{Bekenstein:1973ur} and given a quantum-mechanical foundation by Hawking's discovery of black hole radiation~\cite{Hawking:1975vcx}, remains a central probe of the interface 
	between gravitation and quantum field theory. In $(3+1)$ dimensions, the 
	semiclassical entropy of a Schwarzschild black hole is given by the 
	Bekenstein--Hawking formula,
	\begin{equation}
		S_{\mathrm{BH}} = \frac{\mathcal{A}_{\Sigma}}{4G},
	\end{equation}
	where $\mathcal{A}_{\Sigma}$ is the horizon area. The area-law scaling, as opposed to the volume scaling typical of extensive thermodynamic systems, suggests a geometric, horizon-localized origin for the underlying degrees of freedom.
	A compelling interpretation, introduced by Bombelli, Koul, Lee and Sorkin (BKLS)~\cite{Bombelli:1986} and further developed by Srednicki~\cite{Srednicki:1993}, identifies this entropy with the entanglement between field degrees of freedom inside and outside the horizon. For free Gaussian fields the entanglement entropy is well understood~\cite{Casini:2009, Sugishita:2016, Callan:1994}, and its leading ultraviolet (UV) divergence is the area-law. Susskind and Uglum~\cite{Susskind:1994} argued that the leading area-law divergence is absorbed into the renormalization of Newton's constant, so that the Bekenstein--Hawking entropy, when expressed in terms of the renormalized gravitational coupling, is finite and scheme-independent~\cite{Fursaev:2016, Susskind:1995, Solodukhin:2011}. This identification is central to the geometric-entropy programme.
	
	Extending this picture to interacting quantum fields is essential 
	if the Bekenstein--Hawking entanglement correspondence is to be a generic 
	feature of quantum field theory rather than a property of Gaussian states 
	alone. The foundational formalism, the replica trick on conical 
	manifolds together with the heat-kernel expansion in the presence of 
	conical defects, was established by Fursaev, Solodukhin, and 	Dowker~\cite{Fursaev:2016, Solodukhin:2011, Fursaev:1997, Dowker:1994, 
	Fursaev_Solodukhin:1995}. Within this framework, explicit perturbative 
	computations in flat backgrounds have been carried out for self-interacting scalars. Hertzberg and	Wilczek~\cite{HertzbergWilczek:2011} identified 
	calculable finite contributions to the free-scalar entanglement entropy 
	and Hertzberg~\cite{Hertzberg:2013} subsequently extended the analysis 
	to $\lambda\phi^4$ and $g\phi^3$ theories up to two-loop order, 
	demonstrating that volume divergences cancel and that the leading 
	effect of the interaction is the replacement of the bare mass by the 
	renormalized mass in the free-field area law. 
	
	More recently, Pang and Chen~\cite{PangChen:2021} studied the renormalization of entanglement entropy for $\lambda\phi^4$ on $C_\epsilon \times \mathbb{R}^2$ at order $\lambda$ and carried an explicit computation of the entanglement entropy using a position-space conical-propagator method in dimensional regularization, showing that the theory is renormalizable once all relevant operators of dimension $\leq 4$ consistent with the symmetries 
	of the conical geometry are included, and explicitly mapping the resulting renormalization of the two-dimensional brane-tension counterterm onto the  $\mathcal{A}_\Sigma/4G_R$ form in the infinitely massive (flat-horizon) limit.
	
	The present work builds on and extends this line of investigation in 
	three specific directions. First, we carry out the full first-order 
	computation on the Euclidean Schwarzschild background, treating the non-minimal curvature coupling $\xi R \phi^2$ as a free physical parameter. Because the distributional curvature at the conical tip, $R_{\mathrm{sing}} = 4\pi(1-n)\delta_\Sigma$, couples directly to $\xi$ through the singular heat-kernel coefficient $\delta a_1 = (1/6 - \xi)(n^2-1)/n$, the non-minimal coupling plays a direct and non-trivial role in the horizon entanglement at 
	$\mathcal{O}(\alpha)$ which to our knowledge has not been addressed explicitly. In particular, the first-order 
	correction vanishes identically for a conformally coupled scalar 
	$(\xi = 1/6)$, reflecting the direct coupling of $\xi R$ to the distributional curvature at the tip through the singular heat-kernel coefficient $\delta a_1 \propto (1/6 - \xi)$. Second, working in a proper-time regularization scheme, we identify a mixed ultraviolet divergence of the form $\epsilon^{-2}\ln(m^2\epsilon^2)$ arising from the interference 	between bulk vacuum fluctuations and the conical defect, a structure 	specific to proper-time regularization and absent in dimensional schemes. We show that this log-enhanced quadratic divergence is 
	cancelled exactly, at this order, by the standard mass-renormalization 
	counterterm generated by the tadpole self-energy, leaving only the 
	standard $m^2\ln(m^2\epsilon^2)$ divergence that renormalizes Newton's 
	constant. Third, we obtain a closed-form analytic expression for 
	$\delta S^{(1)}$ as a function of the field mass $m$, the quartic 
	coupling $\alpha$, and the non-minimal coupling $\xi$, and specialize 
	it to the Schwarzschild horizon with 
	$\mathcal{A}_\Sigma = 16\pi G^2 M^2$. This allows us to map the 
	parameter space of the correction: the massless decoupling at fixed 
	$\xi = 0$, the conformal cancellation at $\xi = 1/6$, and the change 
	of sign of $\delta S^{(1)}$ across this value.
	
	Taken together, these results extend the Susskind–Uglum renormalization paradigm to interacting scalars with non-minimal curvature coupling on a curved background, and provide a complementary check, via heat kernels and proper-time regularization, to the dimensional-regularization analysis of Pang and Chen. The structural form $S_{\mathrm{BH}} = \mathcal{A}_\Sigma / 4 G_F$ 	survives the inclusion of quartic self-interactions at $\mathcal{O}(\alpha)$, with the interaction-induced radiative corrections absorbed entirely into a renormalized Newton's constant $G_F$ whose dependence on $(m, \alpha, \xi)$ is determined explicitly.
	
	The paper is organized as follows. Section~\ref{formalism} sets up 
	the Schwarzschild geometry and the replica construction, and reviews 
	the heat-kernel decomposition on manifolds with a conical defect. 
	Section~\ref{Quartic_Interaction} develops the perturbative expansion for the quartic interaction and isolates the regular and singular contributions to the coincident propagator. Section~\ref{Four_Dim} specializes to $d=4$, performs the small-$\epsilon$ expansion, and derives the bare first-order entropy correction. Section~\ref{Renormalization} carries out the matching to the Einstein--Hilbert action and defines the renormalized Newton's 
	constant $G_F$. Appendix~A presents the mass-counterterm cancellation 
	of the mixed divergence in detail. Throughout we work in Euclidean 
	signature with units $\hbar = c = k_B = 1$ and metric signature 
	$(+,+,+,+)$.

	\section{Formalism and Background Geometry}
	\label{formalism}
	
	\hspace{1em} We consider a real, massive, non-minimally coupled scalar field $\phi$ on the four-dimensional Euclidean Schwarzschild background. The Euclidean line element is
	\begin{equation}
		ds^2 = \left(1-\frac{2GM}{r}\right) d\tau^2 + \left(1-\frac{2GM}{r}\right)^{-1} dr^2 + r^2 d\Omega_2^2,
		\label{eq:metric}
	\end{equation}
	with Euclidean time identified periodically, $\tau \sim \tau + \beta_H$, where $\beta_H = 8\pi GM$ is the inverse Hawking temperature. The horizon is a two-sphere $\Sigma$ located at $r_h = 2GM$. The smooth background is Ricci-flat, $R_{\mathrm{smooth}} = 0$, a property that simplifies the heat-kernel coefficients below.
	
	The entanglement entropy $S$ is computed using the standard replica trick~\cite{Callan:1994, Fursaev:2016}. We introduce a conical singularity at the horizon by considering the $n$-fold covering space $\mathcal{M}_n$. In the near-horizon limit, the metric factorizes locally into $C_n \times \Sigma$, where $C_n$ is a two-dimensional cone with deficit angle $2\pi(1-n)$~\cite{Dowker:1994}. The entropy is recovered from the partition function $Z_n$ on this singular manifold via the relation
	\begin{equation}
		S = - \left. \frac{\partial}{\partial n} \left( \ln Z_n - n \ln Z_1 \right) \right|_{n=1}.
		\label{eq:replica_limit}
	\end{equation}
	For a free scalar field, the action on $\mathcal{M}_n$ is quadratic,
	\begin{equation}
		S_E[\phi] = \frac{1}{2} \int_{\mathcal{M}_n} \mathrm{d}^4 x \sqrt{g} \, \phi \mathcal{O}_n \phi,
	\end{equation}
	where the fluctuation operator is given by
	\begin{equation}
		\mathcal{O}_n = -\nabla^2 + m^2 + \xi R.
	\end{equation}
	Here, $\xi$ is the non-minimal coupling to the Ricci scalar $R$. Although the smooth Schwarzschild geometry is Ricci-flat, the replica construction introduces a distributional curvature localized at the tip of the cone, $R_{\text{sing}} = 4\pi(1-n)\delta_\Sigma$, which enters the singular heat-kernel coefficient $\delta a_1$ through its coupling to $\xi$, and is responsible for the interaction-induced entropy correction derived below~\cite{Fursaev_Solodukhin:1995}.
	
	The one-loop effective action is formally defined by the functional determinant, $W_n = -\ln Z_n = \frac{1}{2} \ln \det \mathcal{O}_n$. In the presence of interactions, this free energy receives perturbative corrections, which we compute using the heat kernel method on the conical background~\cite{Vassilevich:2003}.
	
	\section{Quartic Interaction}
	\label{Quartic_Interaction}
	
	\hspace{1em}We deform the Gaussian theory by a weak quartic coupling $\alpha$, treated perturbatively. 
	The Euclidean action on the replicated manifold $\mathcal{M}_n$ is
	\begin{equation}
		S[\phi] = \frac{1}{2}\int_{\mathcal{M}_n} \mathrm{d}^d x \,\sqrt{g}\;\phi\,\mathcal{O}_n\phi 
		+ \frac{\alpha}{4!}\int_{\mathcal{M}_n} \mathrm{d}^d x \,\sqrt{g}\;\phi^4 ,
		\label{eq:interacting_action}
	\end{equation}
	where $\alpha$ is the quartic coupling.
	It is convenient to separate the action as
	\begin{equation}
		S[\phi] = S^{(0)}[\phi] + S_\mathrm{int}[\phi],
	\end{equation}
	with $S^{(0)}$ the Gaussian part and $ S_{\mathrm{int}} = \frac{\alpha}{4!}\int_{\mathcal{M}_n} \mathrm{d}^d x \,\sqrt{g}\;\phi^4 ,$ the interaction term.
	The corresponding partition function factorizes as
	\begin{equation}
		Z_n = Z_n^{(0)} \,\Big\langle e^{-S_{\mathrm{int}}[\phi]} \Big\rangle_{0,n}
	\end{equation}
	where $Z_n^{(0)}$ is the free partition function and 
	$\langle \cdot \rangle_{0,n}$ denotes expectation values in the Gaussian theory.
	
	Expanding for small $\alpha$, the logarithm of the partition function is
	\begin{equation}
		\ln Z_n 
		= \ln Z_n^{(0)} - \langle S_{\mathrm{int}}\rangle_{0,n} 
		+ \mathcal{O}(\alpha^2),
		\label{eq:perturbative_expansion}
	\end{equation}
	which is the first-order perturbative correction.

	Wick’s theorem applies on \(\mathcal{M}_n\) since the combinatorial structure of field contractions is independent of the underlying manifold~\cite{Wick:1950}. For four identical fields at coincident points there are three distinct pairings, yielding
	\begin{equation}
		\ln Z_n = \ln Z_n^{(0)} - \frac{\alpha}{8} \int_{\mathcal{M}_n} \mathrm{d}^d x\, \sqrt{g}\, [G_n(x,x)]^2 + \mathcal{O}(\alpha^2).
		\label{eq:logarithm_partition_function}
	\end{equation}
	
	The coincident propagator \(G_n(x,x)\) is ultraviolet divergent, and consequently the quantity \(G_n(x,x)^2\) requires regularization. To control this divergence we introduce a proper-time regulator \(\epsilon^2\), with the limit \(\epsilon^2 \to 0\) taken at the end of the calculation. The propagator on the \(n\)-sheeted manifold can be represented in terms of the heat kernel as~\cite{Vassilevich:2003}
	\begin{equation}
		G_n(x,x) = \int_{\epsilon^2}^{\infty} ds \, K_{{\mathcal{M}}_n}(x,x;s) \, e^{-m^2 s},
		\label{eq:propagator_heat_kernel}
	\end{equation}
	
	Here \(s\) denotes the diffusion time. Since the manifold \(\mathcal{M}_n\) possesses a conical singularity, the standard heat-kernel expansion valid for smooth manifolds cannot be applied directly; instead, additional singular contributions localized at the tip must be taken into account. 
	
	It was shown that a manifold with a conical singularity can be decomposed into a smooth contribution and a singular part~\cite{Fursaev:1997}.
	
	\begin{equation}
		K_{\mathcal{M}_n}(x;s) = K_{\mathrm{reg}}(x;s) + \delta K_{\mathrm{sing}}(x;s).
	\end{equation}
	
	In the vicinity of the horizon \(\Sigma\), the geometry of \(\mathcal{M}_n\) factorizes as $\mathcal{M}_n \simeq C_n \times \Sigma$~\cite{Fursaev_Solodukhin:1995},
	where \(C_n\) is a two-dimensional cone and \(\Sigma\) is the horizon surface. In $d=4$, only the coefficients $a_0$ and $a_1$ produce UV-divergent contributions to the coincident propagator; $a_k$ with $k \geq 2$ yield finite pieces which do not affect the divergence structure analyzed below.
	
	The regular part satisfies:
	\begin{equation}
		K_{\mathrm{reg}}(x;s) = \frac{1}{(4\pi s)^{d/2}} \left[ a_0^{\mathrm{reg}}(x;n) + a_1^{\mathrm{reg}}(x;n)s + \mathcal{O}(s^2) \right],
	\end{equation}
	with coefficients~\cite{Vassilevich:2003}:
	
	\begin{align}
		a_0^{\mathrm{reg}}(x;n) &= 1,
		\nonumber
		\\
		a_1^{\mathrm{reg}}(x;n) &= \left(\frac{1}{6} - \xi\right) R_{\mathrm{smooth}}(x;n).
		\label{eq:regular_hk_coefficients}
	\end{align}
	
	The singular part is supported on the horizon $\Sigma$:
	\begin{equation}
		\delta K_{\mathrm{sing}}(x;s) = \frac{1}{(4\pi s)^{d/2}} \left[ \delta a_0(x;n) + \delta a_1(x;n)s + \mathcal{O}(s^2) \right] \delta_\Sigma,
	\end{equation}
	
	Consequently, the leading singular coefficients are~\cite{Solodukhin:2011}:
	\begin{align}
		\delta a_0(x;n) &= 0, \\
		\delta a_1(x;n) &= \left(\frac{1}{6} - \xi\right) \left(\frac{n^2 - 1}{n}\right).
	\end{align}
	
	Therefore, Eq.~\eqref{eq:propagator_heat_kernel} can be written as
	\begin{equation}
		G_n(x,x) = 
		\underbrace{\int_{\epsilon^2}^{\infty} ds \, K_\mathrm{reg}(x;s) \, e^{-m^2 s}}_{G_\mathrm{reg}(x,x)} 
		+ 
		\underbrace{\int_{\epsilon^2}^{\infty} ds \, \delta K_\mathrm{sing}(x;s) \, e^{-m^2 s}}_{G_\mathrm{sing}(x,x)},
	\end{equation}
	For our geometry, away from the conical singularity, the Ricci scalar vanishes, $R_{\mathrm{smooth}}(x;n) = 0$. Consequently, the heat kernel coefficient $a_1^\mathrm{reg}$ vanishes, and the leading UV divergence of the regular propagator is determined solely by the universal coefficient $a_0^\mathrm{reg}$ which is position independent, yielding a constant value across the manifold $\mathcal{M}_n$. In contrast, $G_\mathrm{sing}$ is localized entirely on the horizon as indicated by the delta function $\delta_{\Sigma}$. To first order in $s$, the regular and singular parts of the propagator are given by

	\begin{align}
		G_\mathrm{reg}(x,x) &= \int_{\epsilon^2}^{\infty} ds \, 
		\frac{e^{-m^2 s}}{(4 \pi s)^{d/2}},  \\
		G_\mathrm{sing}(x,x) &= \left(\frac{1}{6} - \xi\right) \left( \frac{n^2 - 1}{n} \right) \delta_{\Sigma} \int_{\epsilon^2}^{\infty} ds \, 
		\frac{e^{-m^2 s}}{(4 \pi s)^{d/2}} s
		\label{eq:g_reg_g_sing}.
	\end{align}
	Splitting the propagator as $G_n = G_{\mathrm{reg}} + G_{\mathrm{sing}}$, the coincident square decomposes as
	
	\begin{equation}
		\int_{\mathcal{M}_n} [G_n(x,x)]^2 \;=\; \int_{\mathcal{M}_n} G_{\mathrm{reg}}^2
		\;+\; 2 \int_{\mathcal{M}_n} G_{\mathrm{reg}} G_{\mathrm{sing}}
		\;+\; \int_{\mathcal{M}_n} G_{\mathrm{sing}}^2.
		\label{eq:gn_squared}
	\end{equation}
	However, the crucial point is that the singular heat-kernel coefficient is
	\begin{equation}
		\delta a_1(n) \;=\; \left(\frac{1}{6} - \xi\right)\frac{n^2-1}{n}
		= \left(\frac{1}{3} - 2\xi\right)(n-1) + \mathcal{O}((n-1)^2).
	\end{equation}

	The squared propagator $G_\mathrm{sing}$ formally involves the ill-defined distribution $\delta_{\Sigma}^2$. However, this term scales as $\mathcal{O}((n-1)^2)$. Because $\delta a_1$ vanishes at $n = 1$, the $G^2_{\mathrm{sing}}$ contribution to $\ln Z - n \ln Z_1$ is $\mathcal{O}((n-1)^2)$ and therefore does not contribute to $S = - \left. \frac{\partial}{\partial n} \left( \ln Z_n - n \ln Z_1 \right) \right|_{n=1}$. The regularization associated with $\delta^2_{\Sigma}$ thus drops out of the physical entropy at $\mathcal{O}(\alpha)$~\cite{Solodukhin:2011}.
	
	\section{Four-Dimensional Case}
	\label{Four_Dim}
	\hspace{1em}We now specialize the general results to \(d=4\), corresponding to the physical spacetime. Writing the expressions for $G_{\rm reg}$ and $G_{\rm sing}$ in terms of the incomplete gamma function we obtain
	\begin{align}
		G_{\rm reg}(x,x) 
		&= \frac{m^2}{(4\pi)^2}\Gamma\Big(-1, m^2 \epsilon^2\Big), 
		\nonumber
		\\
		G_{\rm sing}(x,x) 
		&= \frac{1}{(4\pi)^2} \left(\frac{1}{6} - \xi\right) \left( \frac{n^2 - 1}{n} \right) \, \delta_\Sigma(x) \, \Gamma\Big(0, m^2 \epsilon^2 \Big)
		\label{eq:g_reg_g_sing_4D},
	\end{align}
	where \(\Gamma\) denotes the incomplete gamma function.
	The cross-term in Eq.~\eqref{eq:gn_squared} includes the contribution from the conical singularity. The singular part of the propagator is supported on the entangling surface $\Sigma$, represented by a delta-function $\delta_\Sigma(x)$. In the limit $\epsilon^2 \to 0$, $G_{\rm sing}$ becomes an integral over $\Sigma$. Since $G_\mathrm{sing}$ is proportional to $\delta_{\Sigma}$, integration over $\mathcal{M}_n$ reduces the integral over the horizon surface, yielding an explicit factor of $\mathcal{A}_{\Sigma}$.
	
	\begin{align}
		\int_{\mathcal{M}_n} d^4 x \, \sqrt{g} \, G_{\rm reg}(x,x) \, G_{\rm sing}(x,x) 
		&= \frac{1}{16 \pi^2} \left(\frac{1}{6} - \xi\right) \left( \frac{n^2-1}{n} \right) \int_{\mathcal{M}_n} d^4 x \, \sqrt{g} \, G_{\rm reg}(x,x) 
		\left[  \delta_\Sigma(x) \int_{\epsilon^2}^{\infty} ds \, s^{-1} e^{-m^2 s} \right] 
		\nonumber
		\\
		&= \frac{1}{16 \pi^2} \left(\frac{1}{6} - \xi\right) \left( \frac{n^2-1}{n} \right) \int_{\epsilon^2}^{\infty} ds \, s^{-1} e^{-m^2 s} 
		\int_\Sigma d^{2} y \, \sqrt{g_\Sigma} \, G_{\rm reg}(r=0, y)
		\nonumber
		\\
		&\approx \frac{1}{16 \pi^2} \left(\frac{1}{6} - \xi\right) \left( \frac{n^2-1}{n} \right)
		\mathcal{A}_{\Sigma} \, G_{\rm reg}(\Sigma) \, \Gamma\Big(0, m^2 \epsilon^2 \Big)
		\label{eq:cross_product_g},
	\end{align}
	where $g_{\Sigma}$ denotes the induced metric on the surface $\Sigma$ 
	and $\mathcal{A}_{\Sigma}$ represents its area. By the heat kernel decomposition~\cite{Fursaev_Solodukhin:1995}, $G_\mathrm{reg}$ is smooth across $\mathcal{M}_n$, while the conical singularity is entirely contained in $G_\mathrm{sing}$. At leading order in the heat kernel expansion (Eq.~\ref{eq:regular_hk_coefficients}), with $R_{\rm smooth} = 0$ for the Schwarzschild background, only the coefficient $a_0^\mathrm{reg} = 1$ contributes, yielding a position-independent leading divergence for $G_\mathrm{reg}$. Therefore, the factorization in the final line follows immediately, 
	and the delta function integral yields $\mathcal{A}_\Sigma$.
	
	For $\epsilon^2 \ll 1$, Eqs.~\eqref{eq:g_reg_g_sing} and \eqref{eq:cross_product_g}
	can be expanded as
	\begin{align}
		G_{\rm reg} &\approx \frac{m^2}{16 \pi^2} 
		\Biggl(
		\chi + \frac{1}{m^2 \epsilon^2} - 1
		\Biggr),
		\nonumber
		\\[2mm]
		\int_{\mathcal{M}_n} G_{\rm reg} G_{\rm sing} 
		&\approx \frac{\mathcal{A}_\Sigma}{256 \pi^4} \left(\frac{1}{6}-\xi\right) \left(\frac{n^2-1}{n}\right) \,
		\Biggl[
		- \frac{\chi}{\epsilon^2} + m^2
		+ m^2 \chi
		\Bigl( 1 - \chi \Bigr)
		\Biggr],
		\label{eq:propagators_expanded}
	\end{align}
	where $\gamma_E$ is the Euler-Mascheroni constant and $\chi \equiv \gamma_E + \ln(m^2 \epsilon^2)$. The finite constant $m^2$ in the bracket is not a subleading term but a genuine $\mathcal{O}(\epsilon^0)$ contribution at first order in $\alpha$, arising from the product of the leading $1/\epsilon^2$ pole of $G_{\mathrm{reg}}$ with the subleading $m^2\epsilon^2$ piece of $\Gamma(0,m^2\epsilon^2)$. Although it cancels in the counterterm subtraction performed in Appendix~A, retaining it is essential for the consistency of the cancellation at $\mathcal{O}(\alpha)$.
	Using these expansions, the logarithm of the partition function to first order in \(\alpha\) becomes
	
	\begin{align}
		\ln Z_n = \ln Z_n^{(0)} - \frac{\alpha m^4 \Delta^2}{2048 \pi^4}\mathcal{V}({\mathcal{M}_n}) \, -  \frac{\alpha \, \mathcal{A}_\Sigma}{1024 \pi^4} \left(\frac{1}{6}-\xi\right) \left(\frac{n^2-1}{n}\right) \, 
		\Biggl[
		- \frac{\chi}{\epsilon^2} + m^2
		+ m^2 \chi
		\Bigl( 1 - \chi \Bigr)
		\Biggr]
		\label{eq:lnZn_computed}.
	\end{align}
	where $\Delta \equiv \chi + \frac{1}{m^2 \epsilon^2} - 1$ and $\mathcal{V}({\mathcal{M}_n})$ is the volume of the manifold $\mathcal{M}_n$.
	The black hole entropy can be cast in terms of the replica index $n$ as
	\begin{equation}
		S = - \left. \frac{\partial}{\partial n} \Big( \ln Z_n - n \ln Z_1 \Big) \right|_{n=1}.
	\end{equation}
	The leading-order bare quantum correction to the entropy is therefore
	\begin{equation}
		\delta S^{(1)}_\mathrm{bare} \;=\; 
		\frac{\alpha \,\mathcal{A}_\Sigma}{512 \pi^4} \left(\frac{1}{6}-\xi\right) \,
		\Biggl[
		- \frac{\chi}{\epsilon^2} + m^2
		+ m^2 \chi
		\Bigl( 1 - \chi \Bigr)
		\Biggr].
	\end{equation}
	For a Schwarzschild black hole with horizon area 
	$\mathcal{A}_{\Sigma} = 4\pi r_h^2 = 16 \pi G^2 M^2$, this yields
	\begin{equation}
			\delta S^{(1)}_\mathrm{bare} =
			\frac{\alpha G^2 M^2}{32 \pi^3} \left(\frac{1}{6}-\xi\right)
			\Biggl[
			- \frac{\chi}{\epsilon^2} + m^2 + m^2 \chi \Bigl( 1 - \chi \Bigr)
			\Biggr].
	\end{equation}
	
	The first-order correction exhibits a mixed UV divergence proportional to $\epsilon^{-2}\ln(m^2\epsilon^2)$ and a logarithmic divergence which is absorbed into the renormalization of Newton's constant. The mixed divergence arises from the interference between the bulk vacuum fluctuations ($G_{\mathrm{reg}}$) and the geometric response of the conical defect ($G_{\mathrm{sing}}$). However, this divergence is an artifact of the bare perturbation theory. As demonstrated in Appendix~A, this term is cancelled by the standard mass renormalization counterterm required to stabilize the theory.
	
	\section{Renormalization of Newton's constant}
	\label{Renormalization}
	\hspace{1em}The expression in Eq.~\eqref{eq:logarithm_partition_function} is directly related to the quantum effective action of the matter field \cite{Vassilevich:2003},
	\begin{equation}
		W_n \equiv -\ln Z_n \,.	
	\end{equation}
	
	The result for the logarithm of the partition function was computed in Eq~\eqref{eq:lnZn_computed}, by substituting, we obtain the quantum effective action 
	\begin{equation}
		W_n = - \ln Z_n^{(0)} + \frac{\alpha m^4 \Delta^2}{2048 \pi^4}\mathcal{V}({\mathcal{M}_n}) \, + \frac{ \alpha \, \mathcal{A}_\Sigma}{1024 \pi^4} \left(\frac{1}{6}-\xi\right) \left(\frac{n^2-1}{n}\right) \, 
		\Biggl[
		- \frac{\chi}{\epsilon^2} + m^2
		+ m^2 \chi 
		\Bigl( 1 - \chi \Bigr)
		\Biggr].
		\label{eq:quantum_effective_action}
	\end{equation}
	We define the bare one-loop coefficients
	\begin{equation}
		B_0 \equiv B_0^\mathrm{bare} = \frac{\alpha m^4 \Delta^2}{2048 \pi^4}, \hspace{2em}
		B_1 \equiv B_1^\mathrm{bare} = \frac{\alpha}{1024 \pi^4} \left(\frac{1}{6}-\xi\right) \, 
		\Biggl[
		- \frac{\chi}{\epsilon^2} + m^2
		+ m^2 \chi
		\Bigl( 1 - \chi \Bigr)
		\Biggr]
		\label{eq:divergence_contribution}.
	\end{equation}
	
	The bulk volume term contributes to cosmological  constant renormalization, while the surface term contributes to Newton's constant renormalization because it is the term proportional to curvature localized on the entangling surface. Here, $B_1$ represents the bare one-loop contribution to the gravitational coupling. 
	In four dimensions, the Einstein--Hilbert action reads~\cite{Solodukhin:1996}
	\begin{equation}
		S_{EH}[g] = -\frac{1}{16 \pi G_B} \int d^4x \sqrt{g}\, R,
	\end{equation}
	where $G_B$ is the bare Newton's constant.
	
	Comparing with the quantum effective action in Eq.~\eqref{eq:quantum_effective_action}, one identifies $B_0$ as a contribution to the cosmological constant and $B_1$ as a correction to $1/G_B$.  
	
	The total effective action can thus be written as
	
	\begin{equation}
		\Gamma_{\mathrm{total}}[g] = S_{EH}[g] + W_n[g] 
		= -\frac{1}{16 \pi G_B}\int d^4x \sqrt{g}\, R + W_n[g].
	\end{equation}
	
	For the replicated manifold $\mathcal{M}_n$ which has a conical singularity the Einstein-Hilbert term $\int \sqrt{g}R$ is~\cite{Fursaev_Solodukhin:1995}
	\begin{equation}
		\int_{\mathcal{M}_n} \sqrt{g}R = 4 \pi(1-n) \mathcal{A}_{\Sigma} + \int_{\mathcal{M} \backslash \Sigma} \sqrt{g}R.
	\end{equation}
	
	Comparing this to the singular part of the Einstein--Hilbert action, $\frac{1}{16 \pi G_B} \int_{\mathcal{M}_n} \sqrt{g}\, (-R) \supset \frac{n-1}{4 G_B} \, \mathcal{A}_\Sigma$, we see that a term in $W_n$ of the form $C \, (n-1) \, \mathcal{A}_\Sigma$ renormalizes $1/G_B$. Our calculated term, $B_1 \, \frac{n^2-1}{n} \, \mathcal{A}_\Sigma$, is of this form since $\frac{n^2-1}{n} = 2(n-1) + \mathcal{O}((n-1)^2)$.
	The divergence proportional to the Ricci scalar $R$ must be absorbed into a renormalized Newton's constant $G_{\mathrm {F}}$~\cite{Susskind:1994}. The term $B_1$ contains a 'mixed' divergence proportional to $\epsilon^{-2}\ln(m^2\epsilon^2)$. In dimensional regularization, the analogue computation yields no 	$\epsilon^{-2}\ln(m^2\epsilon^2)$ structure because dimensional continuation does not generate a mixed power-log divergence at one loop~\cite{PangChen:2021}; the present cancellation is 
	therefore specific to the proper-time scheme. As shown in Appendix~A, the mass-renormalization counterterm cancels the $\chi / \epsilon^2$ and $m^2 \chi^2$ pieces (along with the finite cross-term constant that appears at the same order), leaving the logarithmic divergence $\propto m^2\chi$ which is absorbed into the renormalized Newton's constant via the scale-$\mu$ subtraction below.

	After the mass-counterterm cancellation, the surviving contribution to the surface coefficient is
	\begin{equation}
		B_1^\mathrm{bare} - B_1^\mathrm{ct} = \frac{\alpha}{1024 \pi^4} \left(\frac{1}{6}-\xi\right) m^2 \chi,
		\label{eq:B1_subtracted}
	\end{equation}
	where $\chi = \gamma_E + \ln(m^2 \epsilon^2)$ still depends on the regulator $\epsilon$. To obtain a regulator-independent renormalization of Newton's constant we introduce a renormalization scale $\mu$ and decompose
	\begin{equation}
		\chi = \ln\!\left(\frac{m^2}{\mu^2}\right) + \bigl[\gamma_E + \ln(\mu^2 \epsilon^2)\bigr].
		\label{eq:chi_split}
	\end{equation}
	The bracketed term depends only on the cutoff and on $\mu$, and is absorbed into the bare Newton's constant through the renormalization condition
	\begin{equation}
		\frac{1}{G_B(\epsilon)} = \frac{1}{G_B(\mu)} - \frac{\alpha}{128\pi^4}\!\left(\frac{1}{6}-\xi\right) m^2 \bigl[\gamma_E + \ln(\mu^2 \epsilon^2)\bigr].
		\label{eq:GB_counterterm}
	\end{equation}
	This defines the renormalized one-loop coefficient at scale $\mu$,
	\begin{equation}
		B_1^\mathrm{ren}(\mu) = \frac{\alpha}{1024 \pi^4} \left(\frac{1}{6}-\xi\right) m^2 \ln\!\left(\frac{m^2}{\mu^2}\right),
		\label{eq:renormalized_B1}
	\end{equation}
	and the renormalized Newton's constant,
	\begin{equation}
		\frac{1}{G_F(\mu)} \equiv \frac{1}{G_B(\mu)} + 8\, B_1^\mathrm{ren}(\mu),
		\label{eq:renormalization_newton}
	\end{equation}
	which is manifestly independent of $\epsilon$. The residual $\mu$-dependence is physical: it encodes the running of $1/G_F$ under the renormalization group, driven at $\mathcal{O}(\alpha)$ by the interaction-induced distortion of the vacuum. Between any two scales,
	\begin{equation}
		\frac{1}{G_F(\mu_2)} - \frac{1}{G_F(\mu_1)} = \frac{\alpha}{128 \pi^4}\!\left(\frac{1}{6}-\xi\right) m^2 \ln\!\left(\frac{\mu_1^2}{\mu_2^2}\right),
	\end{equation}
	which is regulator-independent and carries the physical content of the one-loop correction.
	
	The first-order correction to the entanglement entropy is expressed in terms of $B_1^\mathrm{ren}(\mu)$ as
	\begin{equation}
		\delta S^{(1)}(\mu) = 2\, B_1^\mathrm{ren}(\mu)\, \mathcal{A}_\Sigma,
		\label{eq:dS1_final}
	\end{equation}
	so that the structural form of the Bekenstein--Hawking area law is preserved at $\mathcal{O}(\alpha)$,
	\begin{equation}
		S_{\mathrm{BH}} = \frac{\mathcal{A}_\Sigma}{4\, G_F(\mu)},
		\label{eq:BH_renormalized}
	\end{equation}
	with the classical Newton's constant replaced by the scale-dependent $G_F(\mu)$. At this order, $\delta S^{(1)}$ vanishes identically for a conformally coupled field ($\xi = 1/6$).
	
	\begin{figure*}
		\centering
		\includegraphics[width=0.32\linewidth]{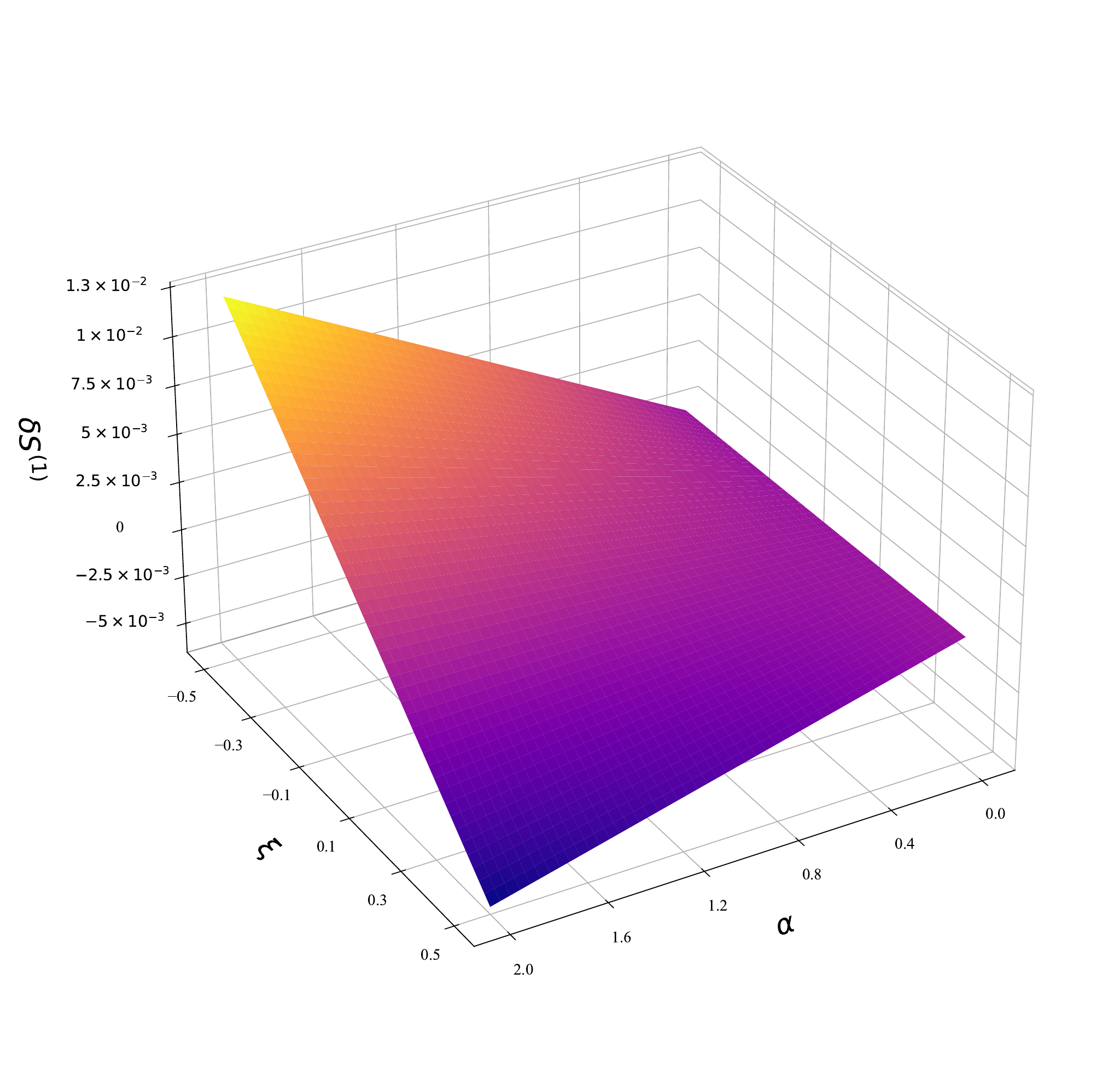}
		\hfill
		\includegraphics[width=0.32\linewidth]{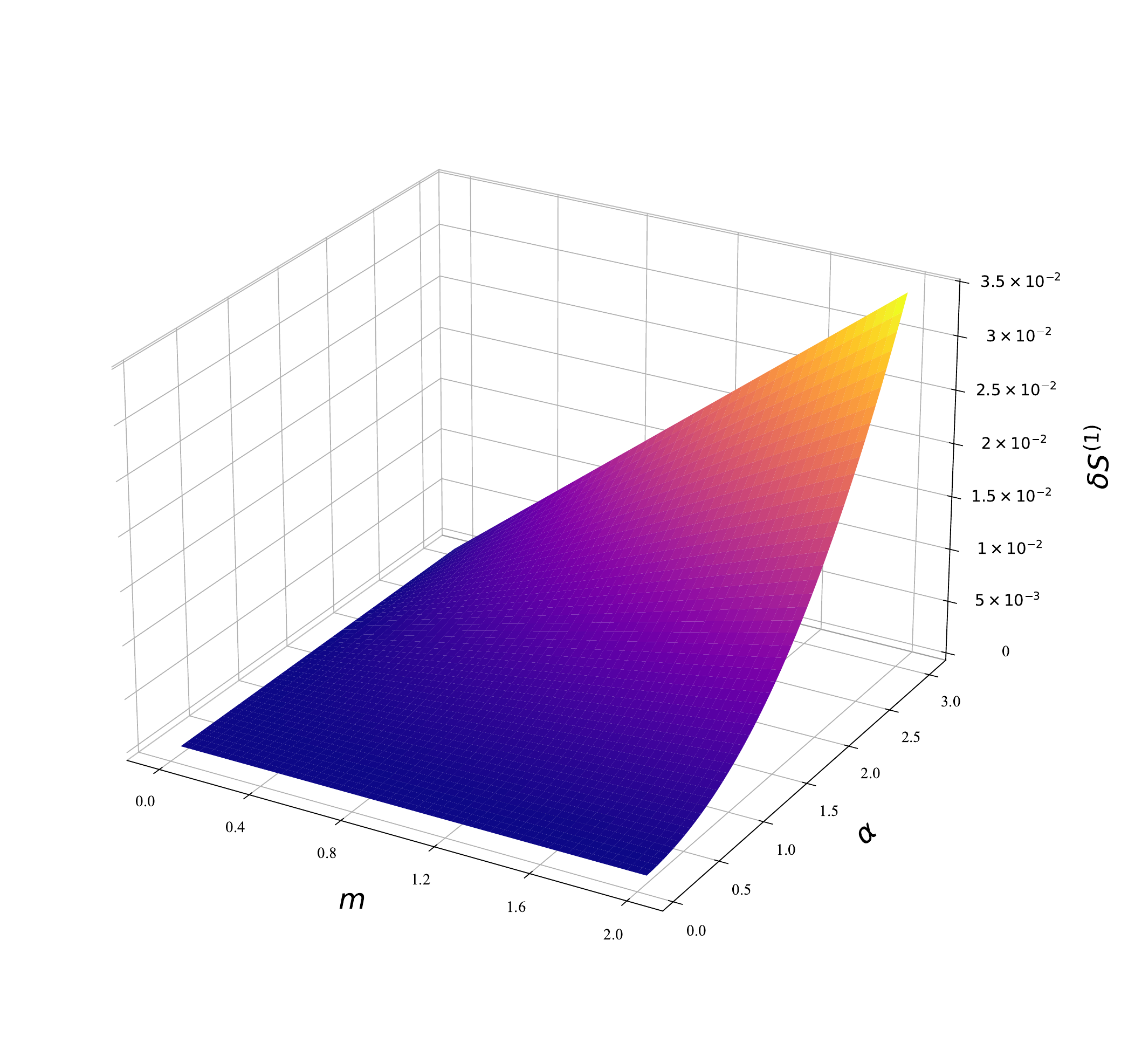}
		\hfill
		\includegraphics[width=0.32\linewidth]{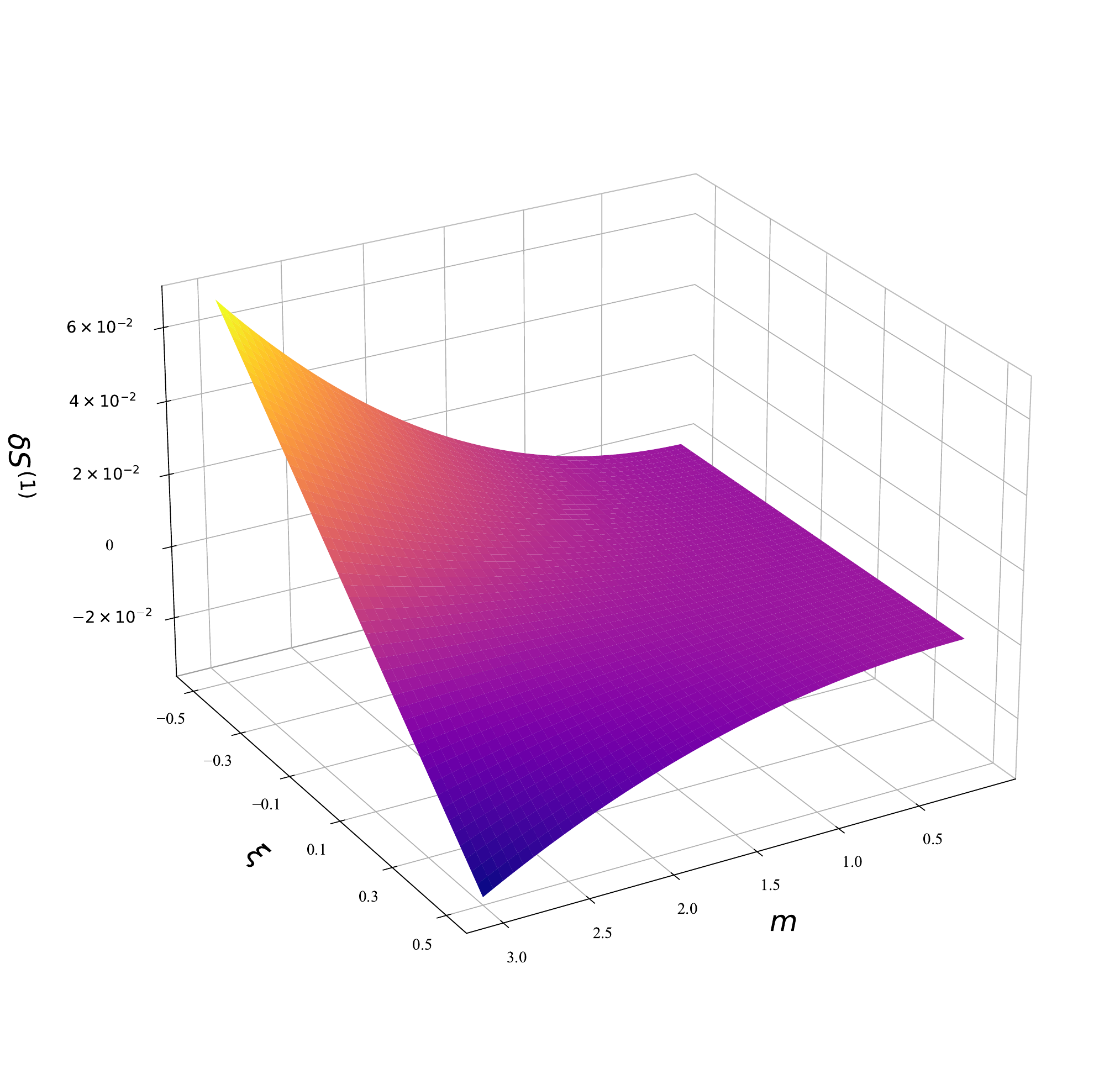}
		\caption{Renormalized first-order entropy correction $\delta S^{(1)}$ 
			to the Bekenstein--Hawking area law, in Planck units with $M = 1$, 
			evaluated at renormalization scale $\mu = 0.01\,M$. 
			\textbf{Left:} $\delta S^{(1)}(\alpha,\xi)$ at fixed $m = 1$. The surface 
			crosses zero at the conformal coupling $\xi = 1/6$, 
			changing sign as $\xi$ crosses this value.
			\textbf{Middle:} $\delta S^{(1)}(\alpha,m)$ at minimal coupling 
			($\xi = 0$). The correction vanishes in the massless limit, reflecting 
			$B_1^{\rm ren} \propto m^2 \ln(m^2/\mu^2) \to 0$.
			\textbf{Right:} $\delta S^{(1)}(m,\xi)$ at fixed $\alpha = 1$. The 
			saddle structure displays the combined non-monotonic dependence on 
			mass and non-minimal coupling, with the zero locus again at 
			$\xi = 1/6$ for $m > \mu$.}
		\label{fig:renormalized_entropy2}
		
	\end{figure*}
	 
	 \section{Conclusions}
	 \hspace{1em}
	 
	 In this work we derived a closed-form analytic expression for the first-order quantum correction 
	 $\delta S^{(1)}$ to the entanglement entropy of a massive, non-minimally coupled self-interacting 
	 scalar field across the horizon of a four-dimensional Schwarzschild black hole. Combining the 
	 replica trick on the conical manifold $\mathcal{M}_n$ with heat-kernel methods in proper-time 
	 regularization, we obtained $\delta S^{(1)}$ as an explicit function of the field mass $m$, the 
	 quartic coupling $\alpha$, the non-minimal coupling $\xi$, and a renormalization scale $\mu$, and 
	 specialized the result to the Schwarzschild horizon with $\mathcal{A}_\Sigma = 16\pi G^2 M^2$.
	 
	 Within the proper-time scheme, the bare correction exhibits divergences of the form $\epsilon^{-2}$ 
	 and $\epsilon^{-2}\ln(m^2\epsilon^2)$, the latter arising from the interference between bulk 
	 vacuum fluctuations and the distributional curvature at the conical tip. This log-enhanced 
	 quadratic divergence is specific to proper-time regularization and does not appear in dimensional 
	 schemes; we showed in Appendix~A that, at $\mathcal{O}(\alpha)$, it is cancelled exactly by the 
	 standard bulk mass-renormalization counterterm once inserted on the replicated manifold. The 
	 residual logarithmic divergence, proportional to $m^2\ln(m^2\epsilon^2)$, is absorbed into a 
	 renormalized Newton's constant defined at scale $\mu$,
	 \begin{equation}
	 	\frac{1}{G_F(\mu)} \equiv \frac{1}{G_B(\mu)} + 8\, B_1^{\rm ren}(\mu),
	 \end{equation}
	 where $B_1^{\rm ren}(\mu) \propto \alpha (1/6 - \xi) m^2 \ln(m^2/\mu^2)$ is manifestly 
	 regulator-independent. The residual $\mu$-dependence encodes the running of $1/G_F$ under the 
	 renormalization group, driven at this order by the interaction-induced distortion of the vacuum. 
	 The structural form of the Bekenstein--Hawking area law is preserved at $\mathcal{O}(\alpha)$,
	 \begin{equation}
	 	S_{\rm BH} = \frac{\mathcal{A}_\Sigma}{4\, G_F(\mu)},
	 \end{equation}
	 with the classical Newton's constant replaced by its scale-dependent counterpart.
	 
	 Taken together, these results extend the Susskind--Uglum renormalization paradigm to interacting  scalars with non-minimal curvature coupling on a curved background, and provide a  methodologically complementary check, via heat kernels and proper-time regularization, to the position-space, dimensional-regularization analysis of Pang and Chen. The explicit  $\xi$-dependence is a distinctive feature: the correction is proportional to $(1/6 - \xi)$ and  vanishes identically at the conformal coupling $\xi = 1/6$ at this order, reflecting the origin 
	 of the singular heat-kernel coefficient $\delta a_1$ in the coupling of $\xi R$ to the  distributional curvature at the tip. For minimal coupling, the interaction generates a non-zero shift in $1/G_F(\mu)$ whose sign depends on the hierarchy between $m$ and $\mu$.
	 
	 Natural directions for future investigation include extension to backgrounds with 
	 $R_{\rm smooth} \neq 0$ (e.g., Reissner--Nordström or de~Sitter), to rotating (Kerr) geometries, 
	 to higher orders in $\alpha$, and to other field species. Whether the $(1/6 - \xi)$ cancellation 
	 at $\mathcal{O}(\alpha)$ survives at two loops, where additional Seeley--DeWitt coefficients 
	 and multiplicative field-strength renormalization enter, is a concrete open question suggested by the present result.
	
	\appendix
	\setcounter{equation}{0}
	\renewcommand{\theequation}{A\arabic{equation}}
	
	\section*{Appendix A}
	To consistently treat ultraviolet divergences, we employ standard Renormalized Perturbation Theory. We define the bare mass as $m_0^2 = m^2 + \delta m^2$ so that the counterterm cancels the divergent part of the one-loop self-energy. The bare Lagrangian is split into a physical part and a counterterm part, introducing the mass counterterm
	\begin{equation}
		\mathcal{L}_{\mathrm{ct}} = \frac{1}{2} \delta m^2 \phi^2.
	\end{equation}
	This interaction generates a first-order correction to the partition function, corresponding to a single counterterm insertion.
	\begin{equation}
		\ln Z_{\mathrm{ct}} = - \frac{1}{2} \int_{\mathcal{M}_n} d^4x \sqrt{g} \, \delta m^2 \, G_n(x,x).
	\end{equation}
	The value of the counterterm $\delta m^2$ is fixed by the renormalization condition that the physical mass of the field remains $m$ \cite{Peskin:1995}. By using the scheme-dependent proper-time cutoff, this requires $\delta m^2$ to cancel the divergent part of the one-loop self-energy in the bulk ($\delta m^2 = -\Sigma_\mathrm{loop}$). The mass counterterm $\delta m^2$ is given by the one-loop self-energy which in a quartic interaction is given by the tadpole diagram. This diagram consists of a single vertex with two legs contracted into a loop, yielding a symmetry factor of $\frac{1}{2}$. Consequently, the counterterm is proportional to the divergent part of the regular Green's function
	\begin{equation}
		\delta m^2 = - \frac{\alpha}{2} [G_{\mathrm{reg}}(x,x)]_{\mathrm{div}}.
	\end{equation}
	Substituting this expression back into the counterterm correction yields
	\begin{equation}\ln Z_{\mathrm{ct}} = +\frac{\alpha}{4} \int_{\mathcal{M}_n} d^4x \sqrt{g} \, [G_{\mathrm{reg}}(x,x)]_{\mathrm{div}} \, \left[ G_{\mathrm{reg}}(x,x) + G_{\mathrm{sing}}(x,x) \right].
		\label{eq:counter_term}
	\end{equation}
	Expanding this product reveals two distinct contributions. The term proportional to $G_{\text{reg}}^2$ represents the renormalization of the bulk vacuum energy (cosmological constant). Since this term scales linearly with the volume of the $n$-sheeted manifold, $\mathcal{V}(\mathcal{M}_n) \approx n \mathcal{V}(\mathcal{M}_1)$, it vanishes upon taking the replica limit $(1 - n \partial_n)$ and therefore does not contribute to the entanglement entropy. However, the cross-term interaction specifically targets the mixed divergence.
	
	From Eq~\eqref{eq:propagators_expanded}, keeping the divergent terms 
	\begin{equation}
		\left[G_\mathrm{reg}\right]_\mathrm{div} = \frac{1}{16\pi^2} 
		\left(
			m^2\chi + \frac{1}{\epsilon^2}
		\right).
	\end{equation}
	By integrating over the whole manifold
	\begin{equation}
		 \int_{\mathcal{M}_n} d^4x \sqrt{g} \, [G_{\mathrm{reg}}]_{\mathrm{div}} \, G_{\mathrm{sing}} = \frac{\mathcal{A}_\Sigma}{256 \pi^4} \left(\frac{1}{6}-\xi\right) \left(\frac{n^2-1}{n}\right) \, 
		 \left(
		 - m^2 \chi^2 - \frac{\chi}{\epsilon^2} + m^2
		 \right),
	\end{equation}
	and the cross-term interaction becomes	
	\begin{equation}
		\mathcal{C}_{\mathrm{cross}} = \frac{\alpha}{4} \int_{\mathcal{M}_n}{d^4x\sqrt{g} \, [G_{\mathrm{reg}}]_{\mathrm{div}} \, G_{\mathrm{sing}}} = \frac{\alpha \mathcal{A}_\Sigma}{1024 \pi^4} \left(\frac{1}{6}-\xi\right) \left(\frac{n^2-1}{n}\right) \, 
		\left(
		- m^2 \chi^2 - \frac{\chi}{\epsilon^2} + m^2 
		\right),
	\end{equation}
	Adding this counterterm contribution to the bare cross-term from Eq.~\eqref{eq:propagators_expanded}
	\begin{equation}
		-\frac{\alpha}{4}\int_{\mathcal{M}_n} G_{\rm reg} G_{\rm sing} 
		= \frac{\alpha \mathcal{A}_\Sigma}{1024 \pi^4} \left(\frac{1}{6}-\xi\right) \left(\frac{n^2-1}{n}\right) \,
		\Biggl[
		\frac{\chi}{\epsilon^2} - m^2
		- m^2 \chi
		\Bigl( 1 - \chi \Bigr)
		\Biggr].
	\end{equation}
	Combining these contributions, one obtains
	\begin{equation}
		\frac{\alpha \mathcal{A}_{\Sigma}}{1024 \pi^4} \left(\frac{1}{6}-\xi\right) \left(\frac{n^2-1}{n}\right) \, 
		\left(
		-m^2\chi^2 - \frac{\chi}{\epsilon^2} + m^2 + \frac{\chi}{\epsilon^2} - m^2 - m^2\chi + m^2\chi^2
		\right),
	\end{equation}
	where $\chi = \gamma_E + \ln(m^2\epsilon^2)$. The total cross contribution is
	\begin{equation}
		- \frac{\alpha \mathcal{A}_{\Sigma}}{1024 \pi^4} \left(\frac{1}{6}-\xi\right) \left(\frac{n^2-1}{n}\right) \, m^2\chi.
	\end{equation}
	
	The counterterm cancels the $-\epsilon^{-2}\ln(m^2\epsilon^2)$ and removes the $\ln^2(m^2\epsilon^2)$ term, leaving only a single logarithmic divergence $\propto m^2 \ln(m^2\epsilon^2)$. At this order no independent surface counterterm is required, the divergence is removed by the standard bulk mass renormalization. The remaining logarithmic divergence $\propto m^2 \ln(m^2\epsilon^2)$ is the standard one-loop divergence that renormalizes Newton's constant $G$.

\end{document}